\documentclass[aps,reprint,onecolumn,prl,showpacs,superscriptaddress,floatfix,notitlepage]{revtex4-1}
\usepackage{amssymb,amsmath,amsfonts} 
\usepackage{epsfig,graphicx}
\usepackage{fancybox, calc}

\newcommand{\ket}[1] {\ensuremath{\left| #1 \right>}}
\newcommand{\bra}[1] {\ensuremath{\left< #1 \right|}}

\newcommand{\braket}[2] {\ensuremath{\left< #1 | #2 \right>}}

\newcommand{\coef}[2] {\ensuremath{\sqrt{\frac{#1}{#2}}}}

\newcommand{\head}{\ensuremath{\ket{\text{head}}_R}}
\newcommand{\tail}{\ensuremath{\ket{\text{tail}}_R}}
\newcommand{\mas}{\ensuremath{\ket{\text{h+t}}_R}}
\newcommand{\menos}{\ensuremath{\ket{\text{h-t}}_R}}

\newcommand{\inicial}{\ensuremath{\ket{\text{init}}}}

\newcommand{\arriba}{\ensuremath{\ket{\uparrow}_S}}
\newcommand{\abajo}{\ensuremath{\ket{\downarrow}_S}}
\newcommand{\izq}{\ensuremath{\ket{\leftarrow}_S}}
\newcommand{\dcha}{\ensuremath{\ket{\rightarrow}_S}}

\newcommand{\okbarra}{\ensuremath{\ket{\overline{\text{ok}}}_{\overline{L}}}}
\newcommand{\failbarra}{\ensuremath{\ket{\overline{\text{fail}}}_{\overline{L}}}}
\newcommand{\ok}{\ensuremath{\ket{\text{ok}}_L}}
\newcommand{\fail}{\ensuremath{\ket{\text{fail}}_L}}

\newcommand{\zmas}{\ensuremath{\ket{+1/2}_L}}
\newcommand{\zmenos}{\ensuremath{\ket{-1/2}_L}}

\newtheorem{myteo}{Assumption} 
\newtheorem{myteo2}{Theorem} 
\newtheorem{hipotesis}{Hypothesis}

\begin{document}
\title{Decoherence allows quantum theory to describe the use of
  itself}

\author{Armando Rela\~{n}o} \affiliation{Departamento de Estructura de
  la Materia, F\'{\i}sica T\'ermica y Electr\'onica, and GISC,
  Universidad Complutense de Madrid, Av. Complutense s/n, 28040
  Madrid, Spain} \email{armando.relano@fis.ucm.es}
\begin{abstract}
  We show that the quantum description of measurement based on
  decoherence fixes the bug in quantum theory discussed in
  [D. Frauchiger and R. Renner, {\em Quantum theory cannot
    consistently describe the use of itself}, Nat. Comm. {\bf 9}, 3711
  (2018)]. Assuming that the outcome of a measurement is determined by
  environment-induced superselection rules, we prove that different
  agents acting on a particular system always reach the same
  conclusions about its actual state.
\end{abstract}

\maketitle

\section{I. Introduction}

In \cite{Renner:18} Frauchiger and Renner propose a Gedankenexperiment
to show that quantum theory is not fully consistent. The setup
consists in an entangled system and a set of fully compatible
measurements, from which four different agents infer contradictory
conclusions. The key point of their argument is that all these
conclusions are obtained from {\em certain} results, free of any
quantum ambiguity:

{\em 'In the argument present here, the agents' conclusions are all
  restricted to supposedly unproblematic ``classical'' cases.'}
\cite{Renner:18}

The goal of this paper is to show that this statement is not
true, at least if ``classical'' states arise from quantum mechanics as
a consequence of environment-induced superselection rules. These rules
are the trademark of the decoherence interpratation of the quantum
origins of the classical world. As it is discussed in \cite{Zurek:03}, a
quantum measurement understood as a perfect correlation between a
system and a measuring apparatus suffers from a number of ambiguities,
which only dissapear after a further interaction with a large
environment ---if this interaction does not occur, the agent
performing the measurement cannot be certain about the real state of
the measured system.

The main conclusion of this paper is that the contradictory
conclusions discussed in \cite{Renner:18} dissapear when the role of
the environment in a quantum measurement is properly taken into
account. In particular, we show that the considered cases only become
``classical'' after the action of the environment, and that this
action removes all the contradictory conclusions.

The paper is organized as follows. In Sec. II we review the
Gedankenexperiment proposed in \cite{Renner:18}. In Sec. IIIA, we
review the consequences of understanding a quantum measurement just as
a perfect correlation between a system and a measuring apparatus; this
discussion is based on \cite{Zurek:03,Zurek:81}. In Sec. IIIB, we
re-interpret the Gedankenexperiment taking into account the
conclusions of Sec. IIIA. In particular, we show that the
contradictory conclusions obtained by the four measuring agents
dissapear due to the role of environment. In Sec. IV we summarize our
main conclusions.

\section{II. The Gedankenexperiment}

This section consists in a review of the Gedankenexperiment proposed
in \cite{Renner:18}. The reader familiarized with it can jump to
section III.

\subsection{A. Description of the setup}

The Gedankenexperiment \cite{Renner:18} starts with an initial state
in which a {\em quantum coin} $R$ is entangled with a $1/2$-spin
$S$. A spanning set of the quantum coin Hilbert space is $\left\{ \head, \tail
\right\}$; for the $1/2$ spin we can use the usual basis, $\left\{
  \arriba, \abajo \right\}$. 

The experiment starts from the following state:
\begin{equation}
\inicial = \coef{1}{3} \head \abajo + \coef{2}{3} \tail \dcha.
\label{eq:inicial}
\end{equation}
From this state, four different agents, $W$, $F$, $\overline{W}$, and
$\overline{F}$, perform different measurements. All these measurements
are represented by unitary operators that correlate different parts of
the system with their apparatus. Relying on the Born rule, they infer
conclusions only from {\em certain} results ---results
with probability $p=1$. These conclusions appear to be contradictory.

To interpret the results of measurements on the initial state,
Eq. (\ref{eq:inicial}), it is useful to rewrite it as different
superpositions of linearly independent vectors, that is, by means of
different orthonormal basis. As is pointed in
\cite{Zurek:81,Zurek:03}, this procedure suffers from what is called
basis ambiguity ---due to the superposition principle, different basis
entail different correlations between the different parts of the
system. This problem is specially important when all the coefficents
of the linear combination are equal \cite{Elby:94}; however, it is not
restricted to this case.  We give here four different possibilities
for the initial state given by Eq. (\ref{eq:inicial}):
\begin{equation}
\inicial_{(1)} = \coef{1}{3} \head \abajo + \coef{1}{3} \tail \abajo + \coef{1}{3} \tail \arriba.
\label{eq:uno}
\end{equation}
(The term in $\head \arriba$ does not show up, because its probability in this state is zero).
\begin{equation}
\inicial_{(2)} = \coef{1}{6} \head \dcha - \coef{1}{6} \head \izq + \coef{2}{3} \tail \dcha.
\label{eq:dos}
\end{equation}
(Again, the probability of the term in $\tail \izq$ is zero).
\begin{equation}
\inicial_{(3)} = \coef{2}{3} \mas \abajo + \coef{1}{6} \mas \arriba - \coef{1}{6} \menos \arriba.
\label{eq:tres}
\end{equation}
(And again, the probability of the term $\menos \abajo$ is zero).
\begin{equation}
\inicial_{(4)} = \coef{3}{4} \mas \dcha - \coef{1}{12} \mas \izq - \coef{1}{12} \menos \dcha - \coef{1}{12} \menos \izq.
\label{eq:cuatro}
\end{equation}

It is worth to remark that all $\inicial_{(1)}$, $\inicial_{(2)}$,
$\inicial_{(3)}$ and $\inicial_{(4)}$ are just different
decompositions of the very same state, $\inicial$. In the equations
above we have used the following notation:
\begin{eqnarray}
\dcha &=& \coef{1}{2} \arriba + \coef{1}{2} \abajo, \\ 
\izq &=& \coef{1}{2} \arriba - \coef{1}{2} \abajo, \\ 
\mas &=& \coef{1}{2} \head + \coef{1}{2} \tail, \\
\menos &=& \coef{1}{2} \head - \coef{1}{2} \tail.
\end{eqnarray}

All the statements that the four agents make in this
Gedankenexperiment are based on different measurements performed on
the initial state, given by Eq. (\ref{eq:inicial}); their results are
easily interpreted relying on
Eqs. (\ref{eq:uno})-(\ref{eq:cuatro}). The procedure is designed to
not perform two incompatible measurements. That is, each agent works
on a different part of the setup, so the wave-function collapse after
each measurement does not interfere with the next one. As a
consequence of this, each agent can infer the conclusions obtained by
the others, just by reasoning from their own measurements.

To structure the interpretation of the Gedankenexperiment, we consider
the following hypothesis for the measuring protocol:

\begin{center} 
\ovalbox{\parbox{\columnwidth-2\fboxsep-2\fboxrule-\shadowsize}{ 
\centering 
\begin{hipotesis}[Measurement procedure] 
  \texttt{} To perform a measurement, an initial state in which the system, $S$, and the apparatus, $A$, are uncorrelated, $\ket{\psi} = \ket{s} \otimes \ket{a}$, is transformed into a correlated state, $\ket{\psi'} = \sum_i c_i \ket{s_i} \otimes \ket{a_i}$, by the action of a unitary operator $U^{M}$. We assume that both $\left\{ \ket{s_i} \right\}$ and $\left\{ \ket{a_i} \right\}$ are linearly independent. Therefore, if the outcome of a measurement is $\ket{a_j}$, then the agent can safely conclude that the system is in the state $\ket{s_j}$.
\end{hipotesis} 
}}
\end{center}

\subsection{B. Development of the experiment}

Equations (\ref{eq:uno})-(\ref{eq:cuatro}) provide four different
possibilities to establish a correlation between the system and the
apparatus. Each of the four agents involved in the Gedankenexperiment
works with one of them. Follows a summary of the main results; more
details are given in \cite{Renner:18}.

{\bf Measurement 1.-} Agent $\overline{F}$ measures the state of the
quantum coin $R$ in the basis $\left\{ \head, \tail \right\}$.  

According to hypothesis 1 above, this statement is based on the
following facts. Agent $\overline{F}$ starts from
Eq. (\ref{eq:dos}). Then, she performs a measurement by means of a
unitary operator that correlates the quantum coin and the apparatus in
the following way
\begin{equation}
\left( c_1 \head + c_2 \tail \right) \otimes \ket{\overline{F}_0} \longrightarrow c_1 \head \ket{\overline{F}_1} + c_2 \tail \ket{\overline{F}_2}.
\label{eq:medida1}
\end{equation}
That is, for any initial state of the coin, $\ket{R} = c_1 \head + c_2
\tail$, the state $\ket{\overline{F}_1}$ of the apparatus becomes
perfectly correlated with $\head$, and the state
$\ket{\overline{F}_2}$ becomes perfectly correlated with $\tail$. This
procedure is perfect if $\braket{\overline{F}_1}{\overline{F}_2}=0$,
but this condition is not necessary to distinguish between the two
possible outcomes. Since the same protocol must be valid for any
initial state, the only constraint for coefficents $c_1$ and $c_2$ is
$\left| c_1 \right|^2 + \left| c_2 \right|^2$=1.

As a consequence of this, the measurement performed by agent
$\overline{F}$ consists in
\begin{equation}
\begin{split}
&\left( \coef{1}{6} \head \dcha - \coef{1}{6} \head \izq + \coef{2}{3} \tail \dcha \right) \otimes \ket{\overline{F}_0} \longrightarrow \\
&\longrightarrow \coef{1}{6} \head \ket{\overline{F}_1} \dcha - \coef{1}{6} \head \ket{\overline{F}_1} \izq + \coef{2}{3} \tail \ket{\overline{F}_2} \dcha.
\end{split}
\end{equation}
Furthermore, the quantum coin together with the agent $\overline{F}$
become the laboratory $\overline{L}$:
\begin{eqnarray}
\head \otimes \ket{\overline{F}_1} &\equiv& \ket{h}_{\overline{L}}, \\
\tail \otimes \ket{\overline{F}_2} &\equiv& \ket{t}_{\overline{L}},
\end{eqnarray}
and therefore the state of the whole system becomes
\begin{equation}
\inicial_{(2)} = \coef{1}{6} \ket{h}_{\overline{L}} \dcha - \coef{1}{6} \ket{h}_{\overline{L}} \izq + \coef{2}{3} \ket{t}_{\overline{L}} \dcha.
\label{eq:dosb}
\end{equation}
The main conclusion obtain from this procedure can be written as follows:

{\bf Statement 1.-} If agent $\overline{F}$ finds her apparatus in the
state $\ket{\overline{F}_2}$, then she can safely conclude that the
quantum coin $R$ is in the state $\tail$. Then, as a consequence of
Eq. (\ref{eq:dosb}), she can also conclude that the spin is in state
$\dcha$, and therefore that agent $W$ is going to obtain $\fail$ in
his measurement (see below for details).

{\bf Measurement 2.-} Agent $F$ measures the state of the spin $S$ in
the basis $\left\{ \arriba, \abajo \right\}$.

Again, according to hypothesis $1$, this statement is based on a
perfect correlation between the apparatus and the spin states. In this
case, agent $F$ starts from Eq. (\ref{eq:uno}). Taking into account
the previous measurement, hers gives rise to the following
correlation:
\begin{equation}
\begin{split}
& \left( \coef{1}{3} \ket{h}_{\overline{L}} \abajo + \coef{1}{3} \ket{t}_{\overline{L}} \abajo + \coef{1}{3} \ket{t}_{\overline{L}} \arriba \right) \otimes \ket{F_0} \longrightarrow \\
&\longrightarrow \coef{1}{3} \ket{h}_{\overline{L}} \abajo \ket{F_1} + \coef{1}{3} \ket{t}_{\overline{L}} \abajo \ket{F_1} + \coef{1}{3} \ket{t}_{\overline{L}} \arriba \ket{F_2}.
\end{split}
\end{equation}
It is worth to note that this measurement is totally independent from
the previous one. 

As it happened with agent $\overline{F}$, agent $F$ becomes entangled with her apparatus, and both together conform the laboratory $L$:
\begin{eqnarray}
\abajo \otimes \ket{F_1} &\equiv& \zmenos, \\
\arriba \otimes \ket{F_2} &\equiv& \zmas.
\end{eqnarray}
Then, the whole system becomes
\begin{equation}
\ket{\text{init}}_{(1)} = \coef{1}{3} \ket{h}_{\overline{L}} \zmenos + \coef{1}{3} \ket{t}_{\overline{L}} \zmenos + \coef{1}{3} \ket{t}_{\overline{L}} \zmas.
\label{eq:statement2}
\end{equation}

The main conclusion obtained by agent $F$ can be written as follows:

{\bf Statement 2.-} If agent $F$ finds her apparatus in state
$\ket{F_2}$, then she can safely conclude that the spin $S$ is in
state $\arriba$. Then, as it is shown in Eq. (\ref{eq:statement2}),
she also is certain that laboratory $\overline{L}$ is in state
$\ket{t}_{\overline{L}}$, and therefore she can safely conclude that
agent $\overline{F}$ has obtained $\tail$ in her measurement. Finally,
according to Statement 1, she can be sure that agent $W$ is going to
obtain $\fail$ in his measurement.
 
{\bf Measurement 3.-} Agent $\overline{W}$ measures the laboratory
$\overline{L}$ in the basis $\left\{ \failbarra, \okbarra \right\}$,
where $\failbarra = \left( \ket{h}_{\overline{L}} +
  \ket{t}_{\overline{L}} \right)/\sqrt{2}$, and $\okbarra = \left(
  \ket{h}_{\overline{L}} - \ket{t}_{\overline{L}} \right)/\sqrt{2}$.

Starting from (\ref{eq:tres}), and taking into account all the
previous results, this measurement implies:
\begin{equation}
\begin{split}
&\left( \coef{2}{3} \failbarra \zmenos + \coef{1}{6} \failbarra \zmas - \coef{1}{6} \okbarra \zmas \right) \otimes \ket{\overline{W}} \longrightarrow \\ &\longrightarrow \coef{2}{3} \failbarra \zmenos \otimes \ket{\overline{W}_1}+ \coef{1}{6} \failbarra \zmas \otimes \ket{\overline{W}_1} - \coef{1}{6} \okbarra \zmas \otimes \ket{\overline{W}_2}.
\end{split}
\label{eq:statement3}
\end{equation}
Again, as the meauserment is not on either the spin $S$ or
the quantum coin $R$ or both, it is fully compatible with the previous
ones. And again, agent $\overline{W}$ becomes entangled with his
apparatus, in the same way that agents $F$ and $\overline{F}$
did. However, since no measurements are done over this new composite
system, we do not introduce a new notation: state $\failbarra$ can be
understood as $\failbarra \otimes \ket{\overline{W}_1}$, and $\okbarra$
  as $\okbarra \otimes \ket{\overline{W}_2}$.

The main conclusion that agent $\overline{W}$ obtains can be
written as follows:

{\bf Statement 3.-} If agent $\overline{W}$ finds his apparatus in
state $\ket{\overline{W}_2}$, then he can safely conclude that
laboratory $\overline{L}$ is in state $\okbarra$. Hence, as a
consequence of Eq. (\ref{eq:statement3}), he can also conclude that
laboratory $L$ is in state $\zmas$. Therefore, from statement 2, agent
$\overline{W}$ knows that agent $F$ has obtained $\arriba$ in her
measurement, and from statement 1, he also knows that agent
$\overline{F}$ has obtained $\tail$. Consequently, agent
$\overline{W}$ can be certain that agent $W$ is going to obtain
$\fail$ in his measurement on laboratory $L$.

The key point in \cite{Renner:18} lays here. As all the agents use the
same theory, and as all the measurements they perform are fully
compatible, they must reach the same conclusion. This conclusion is:

{\em Every time laboratory $\overline{L}$ is in state $\okbarra$, then
  laboratory $L$ is in state $\fail$. Hence, it is not possible to find
  both laboratories in states $\okbarra$ and $\ok$, respectively.}

It is worth to remark that agent $W$ must also obtain the same
conclusion from statements 1, 2 and 3.

{\bf Measurement 4.-} As the final step of the process, agent $W$
measures the laboratory $L$ in the basis $\left\{ \fail, \ok
\right\}$, where $\ok = \left( \zmenos - \zmas \right)/\sqrt{2}$, and
$\fail = \left( \zmenos + \zmas \right)/\sqrt{2}$.

Starting from Eq. (\ref{eq:cuatro}), the result of this final
measurement is

\begin{equation}
\begin{split}
&\left( \coef{3}{4} \failbarra \fail + \coef{1}{12} \failbarra \ok - \coef{1}{12} \okbarra \fail + \coef{1}{12} \okbarra \ok \right) \otimes \ket{W} \rightarrow \\ & \rightarrow \coef{3}{4} \failbarra \fail \otimes \ket{W_1} + \coef{1}{12} \failbarra \ok \otimes \ket{W_2} - \coef{1}{12} \okbarra \fail \otimes \ket{W_1} + \coef{1}{12} \okbarra \ok \otimes \ket{W_2}.
\end{split}
\label{eq:cuatrob}
\end{equation}
Therfore, and despite the previous conclusion that agent $W$ has
obtained from statements 1, 2, and 3, after this measurement he can
conclude that {\em the probability of $\overline{L}$ being in state
  $\okbarra$ and $L$ in state $ok$ is not zero, but $1/12$.} This is
the contradiction discussed in \cite{Renner:18}, from which the
authors of this paper conclude that quantum theory cannot consistently
describe the use of itself:

{\em As Eq. (\ref{eq:cuatrob}) establishes that the probability of
  obtaining $\okbarra \ok$ after a proper measurement is $p=1/12$, and
  as the same theory, used to describe itself, allows us to conclude
  that this very same probability should be $p=0$, the conclusion is
  that quantum theory cannot be used as it is used in statements 1, 2
  and 3. That is, quantum theory cannot consistently describe the use
  of itself.}

In the next section, we will prove that this is a consequence of
hypothesis 1, that is, a consequence of understanding a measurement
just as a perfect correlation between a system and a measuring
apparatus. If we consider that a proper measurement
requires the action of an external environment, as it is discussed in
\cite{Zurek:03,Zurek:81}, quantum theory recovers its ability to speak
about itself. Environmental-induced super-selection rules determining
the real state of the system after a measurement removes all the
contradictions coming from statements 1, 2 and 3.

\section{III. Environment-induced superselection rules}

\subsection{A. The problem of basis ambiguity}

In \cite{Zurek:03,Zurek:81}, W. H. Zurek shows that a perfect
correlation, like the one summarized in hypothesis $1$, is {\em not}
enough to determine the result of a quantum measurement. The reason is
the basis ambiguity due to the superposition principle. To understand
this statement, let us consider a simple measurement in which the
state of the quantum coin $R$ is to be determined. This goal can be
achieved by means of the following unitary operator:
\begin{equation}
U^{RA} = \ket{A_1} \otimes \head \bra{\text{head}}_R + \ket{A_2} \otimes \tail \bra{\text{tail}}_R,
\end{equation}
which establishes a perfect correlation between $\head$ and
$\ket{A_1}$, and between $\tail$ and $\ket{A_2}$. Furthermore, if such
apparatus states verify $\braket{A_1}{A_2}=0$, the measurement
is perfect. Starting from an initial state
\begin{equation}
  \ket{\Psi_0} = \left( \coef{1}{3} \head \otimes \ket{A_1} + \coef{2}{3} \tail \otimes \ket{A_2} \right) \otimes \ket{A_0},
\end{equation}
the final state of the composite system, quantum coin plus apparatus,
is
\begin{equation}
  \ket{\Psi} = \coef{1}{3} \head \otimes \ket{A_1} + \coef{2}{3} \tail \otimes \ket{A_2}.
\label{eq:psi}
\end{equation}

This measurement fulfills the conditions for Hypothesis $1$; indeed,
it is equivalent to the one that the agent $\overline{F}$ performs in
measurement 1. However, {\em the basis ambiguity allows us to rewrite
  (\ref{eq:psi}) in the following way:}
\begin{equation}
\ket{\Psi} = \coef{1}{2} \left( \coef{1}{3} \head + \coef{2}{3} \tail \right) \otimes \ket{A'_1} + \coef{1}{2} \left( \coef{1}{3} \head - \coef{2}{3} \tail \right) \otimes \ket{A'_2}. 
\label{eq:psiprima}
\end{equation}
Note that this is the very same state as the one written in
Eq. (\ref{eq:psi}) ---it is obtained from $\ket{\Psi_0}$ as a consequence of the action of
$U^{RA}$. The new states of the apparatus,  
\begin{eqnarray}
\ket{A_1} &=& \coef{1}{2} \left( \ket{A'_1} + \ket{A'_2} \right), \\
\ket{A_2} &=& \coef{1}{2} \left( \ket{A'_1} - \ket{A'_2} \right),
\end{eqnarray}
also fulfill $\braket{A'_1}{A'_2}=0$, so they also give rise to a
perfect measurement.

Let us reinterpret measurement 1, as described in the previous
section, taking into account this result. Hypothesis $1$ establishes
that a measurement is performed when a perfect correlation between a
system and an apparatus has been settled. But, as both
Eqs. (\ref{eq:psi}) and (\ref{eq:psiprima}) fulfill this requirement,
and both represent the very same state, $\ket{\Psi}$, the action of
the operator $U^{RA}$ is not enough to be sure about the final state
of both the system and the measuring apparatus. Indeed, the only
possible conclusion we can reach is:

{\em Measurement $U^{RA}$ cannot determine the final state of the
  system: if the outcome of the aparatus is ``ONE'', the system can
  either be in state $\head$ or state $\coef{1}{3} \head +
  \coef{2}{3} \tail$; and if the outcome of the apparatus is ``TWO'', the
  system can either be in state $\tail$ or state $\coef{1}{3} \head -
  \coef{2}{3} \tail$.}

Hence, measurement 1, understood as the (only) consequence of
Eq. (\ref{eq:medida1}) seems not enough to support the conclusion
summarized in statement 1. Both $\tail$ and $\coef{1}{3} \head -
\coef{2}{3} \tail$ are fully compatible with the output ``TWO'' of the
measuring apparatus.

But what has really happened? What is the real state of the quantum
coin after the measurement is completed? To which state does the wave
function collapse? We know that experiments provide precise results
---Schr\"odinger cats are always found dead or alive, not in a weird
superposition like $\coef{1}{3} \ket{\text{alive}} - \coef{2}{3}
\ket{\text{dead}}$---, so it is not possible that both possibilities
are true. To answer this question, we introduce the following
assumption:

\begin{center} 
\Ovalbox{\parbox{\columnwidth-2\fboxsep-2\fboxrule-\shadowsize}{ 
\centering 
\begin{myteo}[``Classical'' reality] 
  \texttt{} An event has certainly happened (at a certain time in the
  past) if and only if it is the only explanation for the current
  state of the universe.
\end{myteo} 
}} 
\end{center}

This assumption just reinforces our previous conclusion ---from the
measurement $U^{RA}$, that is, from Eq. (\ref{eq:medida1}), we cannot
make a certain statement about the state of the system. Both $\tail$
and $\coef{1}{3} \head - \coef{2}{3} \tail$ are compatible with the
real state of the universe, given by $\ket{\Psi}$ and the measurement
outcome ``TWO''.

This is why W. H. Zurek establishes that {\em something} else has to
happen before we can make a safe statement about the real state of the
system. The procedure described in Hypothesis $1$ constitutes just a
pre-measurement. The measurement itself requires another action,
perfomed by another unitary operator, to determine the real state of
the system. This action is done by an external (and large)
environment, which becomes correlated with the system and the
apparatus. As is described in \cite{Zurek:03,Zurek:81}, after the
pre-measurement is completed, the system plus the apparatus interacts
with a large environment by means of $U^{\mathcal E}$. Let us suppose
that the result of this interaction is
\begin{equation}
  \ket{\Psi}_{{\mathcal E}} = \coef{1}{3} \head \otimes \ket{A_1} \otimes \ket{{\mathcal E}_1} + \coef{2}{3} \tail \otimes \ket{A_2} \otimes \ket{{\mathcal E}_2},
\label{eq:psi_e}
\end{equation}
with $\braket{{\mathcal E}_1}{{\mathcal E}_2}=0$. Then, this interaction
establishes a perfect correlation between environmental and apparatus
states, in a similar way that the pre-measurement correlates the
system and the apparatus. The main difference between these two
processes is given by the following teorem:

\begin{center} 
\shadowbox{\parbox{\columnwidth-2\fboxsep-2\fboxrule-\shadowsize}{ 
\centering 
\begin{myteo2}[Triorthogonal uniqueness theorem \cite{Elby:94}] 
  \texttt{}
Suppose $\ket{\psi} = \sum_i c_i \ket{A_i} \otimes \ket{B_i} \otimes \ket{C_i}$, where $\{ \ket{A_i} \}$ and $\{ \ket{C_i} \}$ are linearly independent sets of vectors, while $\{ \ket{B_i} \}$ is merely noncollinear. Then there exist no alternative linearly independent sets of vectors $\{ \ket{A'_i} \}$ and $\{ \ket{C'_i} \}$, and no alternative noncollinear set $\{ \ket{B'_i} \}$, such that $\ket{\psi} = \sum_i d_i \ket{A'_i} \otimes \ket{B'_i} \otimes \ket{C'_i}$. (Unless each alternative set of vectors differs only trivially from the set it replaces.) 
\end{myteo2} 
}} 
\end{center}
 
In other words, this theorem establishes that the state
$\ket{\Psi}_{{\mathcal E}}$ is unique, that is, we cannot find another
decomposition for the very same state
\begin{equation}
\ket{\Psi}_{{\mathcal E}} = \coef{1}{2} \left( \coef{1}{3} \head + \coef{2}{3} \tail \right) \otimes \ket{A'_1} \otimes \ket{{\mathcal E}'_1} + \coef{1}{2} \left( \coef{1}{3} \head - \coef{2}{3} \tail \right) \otimes \ket{A'_2} \otimes \ket{{\mathcal E}'_2}, 
\label{eq:psiprima_e}
\end{equation}
with $\braket{{\mathcal E}'_1}{{\mathcal E}'_2}=0$. Hence, the
interaction with the environment determines the real state of the
system plus the apparatus. The action of $U^{{\mathcal E}}$ gives rise
to Eq. (\ref{eq:psi_e}). {\em To obtain a state like the one written
  in Eq. (\ref{eq:psiprima_e}), a different interaction with the
  environment is mandatory, $U^{{\mathcal E}'}$.} Thus, we can
formulate an alternative hypothesis:

\begin{center} 
  \ovalbox{\parbox{\columnwidth-2\fboxsep-2\fboxrule-\shadowsize}{

~

      \centering \begin{hipotesis}[Real measurement procedure (adapted
        from \cite{Zurek:03})]
        \texttt{} To perform a measurement, an initial state in which
        the system, $S$, the apparatus, $A$, and an external
        environment ${\mathcal E}$ are uncorrelated, $\ket{\psi} =
        \ket{s} \otimes \ket{a} \otimes \ket{\varepsilon}$, is
        transformed: {\em i)} first, into a state, $\ket{\psi'} =
        \left( \sum_i c_i \ket{s_i} \otimes \ket{a_i} \right) \otimes
        \ket{\varepsilon}$, by means of a procedure called
        pre-measurement; and {\em ii)} second, into a final state,
        $\ket{\psi''} = \left( \sum_i c_i \ket{s_i} \otimes \ket{a_i}
          \otimes \ket{\varepsilon_i} \right)$. This final state
        determines the real correlations between the system and the
        apparatus. If $\braket{\varepsilon_i}{\varepsilon_j}=0$, then, after tracing out the environmental degrees of freedom, the state becomes
\begin{equation}
\rho = \sum_i \left| c_i \right|^2 \ket{s_i} \ket{a_i} \bra{s_i} \bra{a_i}.
\end{equation}
Therefore, the measuring agent can safely conclude that the result of
the measurement certainly is one of the previous possibilities,
$\left\{ \ket{a_i} \ket{s_i} \right\}$, each one with a probability
given by $p_i = \left| c_i \right|^2$. The states $\left\{ \ket{a_i}
  \ket{s_i} \right\}$ are called ``pointer states''. They are selected
by the environment, by means of environmental-induced superselection
rules; they constitute the ``classical'' reality.

~

\end{hipotesis} 
}}
\end{center}

This hypothesis establishes that only after the real correlations between the system, the apparatus and the
environment are settled, the observation of the agent becomes
certain. Tracing out the environmental degrees of freedom, which are
not the object of the measurement, the state given by
Eq. (\ref{eq:psi_e}) becomes:
\begin{equation}
\rho_{{\mathcal E}} = \frac{1}{3} \head \ket{A_1} \bra{\text{head}}_R \bra{A_1} + \frac{2}{3} \tail \ket{A_2} \bra{\text{tail}}_R \bra{A_2}. 
\end{equation}
In other words, the agent observes a mixture between the system being
in state $\head$ with the apparatus in state $\ket{A_1}$ (with
probability $p_1=1/3$), and the system being in state $\tail$ with the
apparatus in state $\ket{A_2}$ (with probability $p_2=2/3$). And this
happens because the environment has {\em chosen} $\tail$ and $\head$
as the ``classical'' states ---the ones observed as a consequence of
quantum measurements--- by means of environmental-induced
superselection rules. In other words, following this interpretation,
Schr\"odinger cats are always found either dead or alive because the
interaction with the environment determines that $\ket{\text{dead}}$
and $\ket{\text{alive}}$ are the pointer ``classical'' states.

\subsection{B. Re-interpretation of the Gedankenexperiment}

Let us re-interpret the first statement of the Gedankenexperiment,
in the terms discussed above. Agent
$\overline{F}$ cannot reach any conclusion about the real state of the
quantum coin before the pointer states are obtained by means of the
interaction with a large environment ${\mathcal E}$. The key point is
that {\em the environment ${\mathcal E}$ interacts with the whole
  system, that is, with the quantum coin $R$, the apparatus, and the
  spin $S$, because the three of them are entangled}. So, let us
assume that a correlation like Eq. (\ref{eq:psi}) has happened as a
consequence of the pre-measurement. In such a case, taking into
account that the quantum coin $R$ is entangled with the spin $S$, the
state after the pre-measurement is
\begin{equation}
\ket{\Psi} = \coef{1}{3} \head \abajo \ket{A_1} + \coef{2}{3} \tail \dcha \ket{A_2}.
\end{equation}
The next step in the process is the interaction with the environment,
which determines the pointer states of the system composed by the
quantum coin and the spin. There are several possibilities for such an
interaction. Let us consider, for example,
\begin{equation}
U^{{\mathcal E}} = \ket{\varepsilon_1} \head \abajo \ket{A_1} \bra{\text{head}}_R \bra{\downarrow}_S \bra{A_1} + \ket{\varepsilon_2} \tail \dcha \ket{A_2} \bra{\text{tail}}_R \bra{\rightarrow}_S \bra{A_2}, 
\label{eq:environment1}
\end{equation}
and
\begin{equation}
\begin{split}
U^{{\mathcal E}'} &= \ket{\varepsilon'_1} \head \abajo \ket{A_1} \bra{\text{head}}_R \bra{\downarrow}_S \bra{A_1} + \\ &+ \ket{\varepsilon'_2} \tail \abajo \ket{A_2} \bra{\text{tail}}_R \bra{\downarrow}_S \bra{A_2} + \ket{\varepsilon'_3} \tail \arriba \ket{A_2} \bra{\text{tail}}_R \bra{\uparrow}_S \bra{A_2}.
\end{split}
\label{eq:environment2}
\end{equation}

If the real interaction with the environment is given by
Eq. (\ref{eq:environment1}), the final state of the system, after
tracing out the environmental degrees of freedom, is
\begin{equation}
\rho^{{\mathcal E}} = \text{Tr}_{{\mathcal E}} \left[ U^{{\mathcal E}} \ket{\Psi} \right] = \frac{1}{3} \head \abajo \ket{A_1} \bra{\text{head}}_R \bra{\downarrow}_S \bra{A_1} + \frac{2}{3} \tail \dcha \ket{A_2} \bra{\text{tail}}_R \bra{\rightarrow}_S \bra{A_2}.
\label{eq:final1}
\end{equation}
On, the contrary, if the real interaction with the environment is
given by Eq. (\ref{eq:environment2}), the final state is
\begin{equation}
\begin{split}
\rho^{{\mathcal E}'} &= \text{Tr}_{{\mathcal E}} \left[ U^{{\mathcal E}'} \ket{\Psi} \right] = \\ &= \frac{1}{3} \head \abajo \ket{A_1} \bra{\text{head}}_R \bra{\downarrow}_S \bra{A_1} + \frac{1}{3} \tail \abajo \ket{A_2} \bra{\text{tail}}_R \bra{\downarrow}_S \bra{A_2} + \frac{1}{3} \tail \arriba \ket{A_2} \bra{\text{tail}}_R \bra{\uparrow}_S \bra{A_2}.
\end{split}
\label{eq:final2}
\end{equation}

At this point, the question is: what is the real state of the system
after the measurement is completed? 

\begin{itemize}

\item Eq. (\ref{eq:environment1}) establishes that it is a mixture in which
the agent can find the system either in $\head$ and $\abajo$, with
probability $p=1/3$, or in $\tail$ and $\dcha$, with probability
$p=2/3$. It is worth to remark that this is not a quantum
superposition, but a classical mixture. That is, due to the
interaction with the environment, $U^{{\mathcal E}}$, the state of the
system is compatible with either a collapse to $\head \abajo$, with
$p=1/3$, or a collapse to $\tail \dcha$, with $p=2/3$. This is what
agent $\overline{F}$ concludes in statement 1.

\item But the other possible interaction with the environment,
Eq. (\ref{eq:environment2}), establishes that the real state of the
system is a mixture in which the agent can find the system in $\head$
and $\abajo$, with probability $p=1/3$, $\tail$ and $\abajo$, with
probability $p=1/3$, and $\tail$ and $\arriba$, with probability
$p=1/3$.

\end{itemize}

At this stage, the key point is the following. As the measurement
performed by agent $\overline{F}$ only involves the quantum coin $R$,
her apparatus only reads $\head$ with probability $p=1/3$, and
$\tail$, with probability $p=2/3$. {\em But both (\ref{eq:final1}) and
  (\ref{eq:final2}) are compatible with this result}. Hence, following
assumption 1, {\em agent $\overline{F}$ cannot be certain about the
  state of the spin $S$ ---and thus, she can neither be certain about
  what agent $W$ is going to find when he measures the state of the
  laboratory $L$}. The only way to distinguish between
(\ref{eq:final1}) and (\ref{eq:final2}) is to perform a further
measurement on the spin $S$. Such a procedure would provide the
pointer ``classical'' states of the system composed by the quantum
coin and the spin ---its outcome would determine whether the
interaction with the environment is given by
Eq. (\ref{eq:environment1}) or by Eq. (\ref{eq:environment2}). But
such a procedure would be incompatible with the measurement performed
by agent $F$. Hence, agent $\overline{F}$ has to choose between: {\em
  i)} not being certain about the real state of the quantum spin $S$,
and therefore not being able to reach any conclusion about the
measurement that agent $W$ will do in the future; or {\em ii)}
performing a further measurement which would invalidate the
conclusions of this Gedankenexperiment.

This conclusion is enough to rule out the contradictions discussed in
\cite{Renner:18}. As agent $\overline{F}$ cannot be certain about the
outcome what agent $W$ will obtain in his measurement, none of the
four agents can conclude that it is not possible to find laboratory
$L$ in state $\ok$ and laboratory $\overline{L}$ in state $\okbarra$
at the same time. Hence, the outcome of measurement 4, whatever it is,
becomes fully compatible with all the conclusions obtained by all the
agents.

It is worth to note that the same analysis can also be done over
measurements 2 and 3. The conclusions are pretty the same.

\section{IV. Conclusions}

The main result of this paper is to show that assumption 1 and
hypothesis 2 allow quantum theory to consistently describe the use of
itself. This conclusion is based on the decoherence interpretation
about quantum measurements \cite{Zurek:03}. Hence, a
further statement can be set down:

{\em To make quantum theory fully consistent, in order it can be used
  to describe itself, the decoherence interpretation of measurements
  (and origins of the classical world) is mandatory.}

In any case, the main conclusion of this paper is applicable to other
interpretations of quantum mechanics. Decoherence interpretation of
the meausrement process establishes that the wave-function collapse is
not real ---the measuring agent {\em sees} the system as if its
wave-function had collapsed onto one of the pointer states selected by
the environment, even though the whole wave function remains in a
quantum superposition. However, this interpretation is not really
important for experimental results; from this point of view, it is
compatible with the Copenhaguen interpretation, because it assigns the
same probabilities to all of the possibles outcomes. Furthermore, it
is also compatible with Everett many-worlds interpretation
\cite{Everett:57}: the branches onto which the universe splits after a
measurement are determined by the environmental-induced
super-selection rules. The key point is that real ``classical'' states
are not ambiguous, but they are the (unique) result of the interaction
between the measured system, the measuring apparatus, and a large
environment.

\section{Acknowledgements}

The author is supported by Spain’s MINECO/FEDER Grant
No. FIS2015-70856-P, and acknowledges C. M. L\'obez and A. L. Corps
for their vaulable comments.


\begin{thebibliography}{99}

\bibitem{Renner:18} D. Frauchiger and R. Renner, {\em Quantum theory cannot consistently describe the use of itself}, Nat. Comm. {\bf 9}, 3711 (2018).

\bibitem{Zurek:03} W. H. Zurek, {\em Decoherence, einselection, and the quantum origins of the classical}, Rev. Mod. Phys. {\bf 75}, 715 (2003).

\bibitem{Zurek:81} W. H. Zurek, {\em Pointer basis of quantum apparatus: Into what mixture does the wave packet collapse?}, Phys. Rev. D {\bf 24}, 1516 (1981).

\bibitem{Elby:94} A. Elby and J. Bub, {\em Triorthogonal uniqueness theorem and its relevance to the interpretation of quantum mechanics}, Phys. Rev. A {\bf 49}, 4213 (1994).

\bibitem{Everett:57} H. Everett, {\em "Relative State" Formulation of Quantum Mechanics}, Rev. Mod. Phys. {\bf 29}, 454 (1957).

\end{thebibliography}
\end{document}